\documentclass[%
 aip,
 amsmath,amssymb,
preprint,%
fleqn,
]{revtex4-1}

\usepackage{graphicx}
\usepackage{dcolumn}
\usepackage{bm}

\begin{document}


\title{ Higher order stability of dust ion acoustic solitary wave solution described by the KP equation in a collisionless unmagnetized
nonthermal plasma in presence of isothermal positrons}

\author{Sankirtan Sardar}
\affiliation{ Department of Mathematics, Jadavpur University, Kolkata - 700032, India.}%
\author{Anup Bandyopadhyay}
\email{abandyopadhyay1965@gmail.com}%
\affiliation{ Department of Mathematics, Jadavpur University, Kolkata - 700032, India.}%
\author{K. P. Das}
\affiliation{ Department of Applied Mathematics, University of Calcutta, 92 Acharya Prafulla Chandra Road, Kolkata - 700009, India.}%
\begin{abstract}
\noindent Sardar \textit{et al.} [Phys. Plasmas \textbf{23}, 073703 (2016)] have studied the stability of small amplitude dust ion acoustic solitary waves in a collisionless unmagnetized electron - positron - ion - dust plasma. They have derived a Kadomtsev Petviashvili (KP) equation to investigate the lowest - order stability of the solitary wave solution of the Korteweg-de Vries (KdV) equation for long-wavelength plane-wave transverse perturbation when the weak dependence of the spatial coordinates perpendicular to the direction of propagation of the wave is taken into account. In the present paper, we have extended the lowest - order stability analysis of KdV solitons given in the paper of Sardar \textit{et al.} [Phys. Plasmas \textbf{23}, 073703 (2016)] to higher order with the help of multiple-scale perturbation expansion method of Allen and Rowlands [J. Plasma Phys. \textbf{50}, 413 (1993); \textbf{53}, 63 (1995)]. It is found that solitary wave solution of the KdV equation is stable at the order $k^{2}$, where $k$ is the wave number for long-wavelength plane-wave perturbation.
\end{abstract}

\maketitle

\section{Introduction}
The small-$k$ perturbation expansion method of Rowlands and Infeld \cite{rowlands1969stability,infeld1972stability,infeld1973stability,zakharov1974instability,infeld1985self} is generally used to analyse the lowest order stability of solitary wave solutions of different nonlinear evolution equations in plasmas, where $k$ is the wave number for long-wavelength plane-wave perturbation. Several authors \cite{das1989ion,mamun1996stability,chakraborty1998stability,ghosh1998three,bandyopadhyay1999stability,BD00b,chakraborty2003existence,IBD08b,shalaby2009stability,el2011three,kundu2013dust,saini2014zakharov,khaled2014dust,gill2015electrostatic,sardar2016stability,sardar2016existence,sardar2017existence} have used this method to investigate the lowest order stability of solitary waves in plasmas with or without magnetic field. 

The small-$k$ perturbation expansion method of Rowlands and Infeld \cite{rowlands1969stability,infeld1972stability,infeld1973stability,zakharov1974instability,infeld1985self} fails to study the higher order stability of solitary waves in plasmas. This method also fails to investigate the lowest order stability of the double layers in plasmas. Allen and Rowlands \cite{allen1993determination}
developed a method to analyse the higher order stability of solitary wave solution of the Zakharov-Kuznetsov (ZK) equation. Using this method, Allen and Rowlands \cite{allen1995stability} have derived higher order growth rate of instability for obliquely propagating solitary wave solution of the ion acoustic waves in a magnetized plasma. Several authors \cite{BD01a,BD02a,mp04,chakraborty2004higher,pm05,tian2007stability,das2014stability} have used this method to analyse the higher order stability of solitary wave solutions of the different evolution equations. Bandyopadhyay \& Das \cite{BD01a} have used this method of Allen and Rowlands \cite{allen1993determination,allen1995stability} to find the higher order (i.e., of order $k^{2}$) growth rate of instability of solitary wave solutions of Korteweg-de Vries-Zakharov-Kuznetsov (KdV-ZK) equation and modified KdV-ZK (MKdV-ZK) equation. Later, Parkes and Munro \cite{pm05} have pointed out the error appearing in the growth rate of instability up to the order $k^{2}$ of Bandyopadhyay \& Das. \cite{BD01a} Using the same method, Bandyopadhyay \& Das \cite{BD01b} have investigated the lowest order stability of the double layer solution of the combined MKdV - KdV - ZK equation. Das \textit{et al.} \cite{DBD06,das2007alternative, das2007existence} have used the same multiple scale perturbation expansion method to study the lowest order stability of solitary wave solutions of the complicated evolution equations.  Das \textit{et al.} \cite{das2014stability} have used this method of Allen and Rowlands \cite{allen1993determination,allen1995stability} to investigate the higher order stability of solitary wave solution of the Schamel's modified Korteweg-de Vries - Zakharov-Kuznetsov (SKdV - ZK) equation. Higher order stability analysis of the solitary wave solution of the Schamel's modified Kadomtsev Petviashvili (SKP) equation has been discussed by Chakraborty and Das \cite{chakraborty2004higher}. In a later paper, Tian-Jun\cite{tian2007stability} also investigated the higher order stability of the solitary wave solution of the SKP equation with positive and negative dispersion. In the present paper, we have used the same multiple-scale perturbation expansion method of Allen and Rowlands \cite{allen1993determination,allen1995stability} to investigate the higher order stability of solitary wave solution of the Kadomtsev Petviashvili (KP) equation. For this purpose we have considered the KP equation derived by Sardar \textit{et al.} \cite{sardar2016stability} and this KP equation describes the nonlinear behaviour of the dust ion acoustic (DIA) waves in electron - positron - ion - dust (e-p-i-d) plasma.  

Sardar \textit{et al.} \cite{sardar2016stability,sardar2016existence,
sardar2017existence} have used the small-$k$ perturbation expansion method of Rowlands and Infeld \cite{rowlands1969stability,infeld1972stability,infeld1973stability,zakharov1974instability,infeld1985self} to investigate the lowest order stability of DIA solitary wave solutions of different nonlinear evolution equations describing the nonlinear behaviour of DIA waves in a collisionless unmagnetized e-p-i-d plasma consisting of warm adiabatic ions, static negatively charged dust grains, nonthermal electrons and isothermal positrons. In particular, Sardar \textit{et al.} \cite{sardar2016stability} have investigated the stability of the solitary wave solutions of the KdV and different modified KdV equations with the help of KP and different modified KP equations describing the nonlinear behaviour of DIA waves in different region of parameter space when the weak dependence of the spatial coordinates perpendicular to the direction of propagation of the wave is taken into account. In the present paper, we have extended the lowest order stability analysis of KdV solitons given in the  paper of Sardar \textit{et al.} \cite{sardar2016stability} to higher order with the help of multiple-scale perturbation expansion method of Allen and Rowlands. \cite{allen1993determination,allen1995stability}

Starting from the set of basic equations consisting of the equation of continuity of ions, equation of motion of ions, the pressure equation for ion fluid, the Poisson equation, the equation for the number density of the nonthermal electrons of Cairns \textit{et al.} \cite{cairns1995electrostatic}, the equation for the number density of isothermal positrons and the unperturbed charged nutrality condition, Sardar \textit{et al.} \cite{sardar2016stability} have derived the following three-dimensional Kadomtsev Petviashvili (KP) equation:
\begin{eqnarray}\label{KP_equation}
\frac{\partial}{\partial \xi}\Big[\phi^{(1)}_{\tau} + AB_{1}
\phi^{(1)} \phi^{(1)}_{\xi} + \frac{1}{2} A C \phi^{(1)}_{\xi \xi \xi}\Big] + \frac{1}{2} AD \Big( \phi^{(1)}_{\eta \eta } + \phi^{(1)}_{\zeta \zeta}\Big) = 0.
\end{eqnarray}
Here $\phi^{(1)}$ is the first order perturbed electrostatic potential, $\xi$, $\eta$, $\zeta$ are the stretched spatial coordinates and $\tau$ is the stretched time coordinate and we have used the following notations: $\displaystyle \phi^{(1)}_{\tau} = \frac{\partial \phi^{(1)}}{\partial \tau}$, $\displaystyle \phi^{(1)}_{\xi} = \frac{\partial \phi^{(1)}}{\partial \xi}$, $\displaystyle \phi^{(1)}_{\xi\xi\xi} = \frac{\partial^{3} \phi^{(1)}}{\partial \xi^{3}}$, $\displaystyle \phi^{(1)}_{\eta\eta} = \frac{\partial^{2} \phi^{(1)}}{\partial \eta^{2}}$,  $\displaystyle \phi^{(1)}_{\zeta\zeta} = \frac{\partial^{2} \phi^{(1)}}{\partial \zeta^{2}}$. 

The coefficients $A$, $B_{1}$, $C$ and $D$ are, respectively, given in the equations (17), (18), (5) and (19) of Sardar \textit{et al.} \cite{sardar2016stability} They have used the appropriate stretchings of the independent variables and the appropriate perturbation expansions of the dependent variables to derive the three-dimensional KP equation (\ref{KP_equation}). This KP equation admits the following solitary wave solution:
\begin{eqnarray}\label{phi1}
\phi^{(1)}=\phi_{0}(X)=a \mbox{sech}^{2}\Big[\frac{X}{W}\Big] ,
\end{eqnarray}
where $X=\xi-U\tau$ and $U$ is the dimensionless velocity (normalized by $C_{D}$, $C_{D}$ = linearized velocity of the DIA wave for long wave length plane wave pertubation of the present plasma system) of the travelling wave moving along $x$ - axis, i.e., $U$ is the dimensionless velocity of the wave frame. 

The amplitude ($a$) and the width ($W$) of the solitary wave solution (\ref{phi1}) are given by
\begin{eqnarray}\label{a_combined}
a=\frac{3 U} { AB_{1}} \mbox{  and  }W^{2} = \frac{2 A C}{U}.
\end{eqnarray}

The solution (\ref{phi1}) is the steady state solution of the KP equation (\ref{KP_equation}) along the $x$ - axis. This solution is same as the solitary wave solution of the KdV equation corresponding to the KP equation (\ref{KP_equation}), i.e., the steady state solitary wave solution of the KdV equation
\begin{eqnarray}\label{KdV_equation}
\phi^{(1)}_{\tau} + AB_{1} \phi^{(1)} \phi^{(1)}_{\xi} + \frac{1}{2} A C \phi^{(1)}_{\xi \xi \xi}= 0
\end{eqnarray}
is exactly same as the equation (\ref{phi1}). Sardar \textit{et al.} \cite{sardar2016stability} have studied the lowest order transverse stability of the solitary wave solution of the KdV equation (\ref{KdV_equation}) using the three-dimensional KP equation (\ref{KP_equation}). In the present paper, our aim is to study the higher order transverse stability of the solitary wave solution of the KdV equation (\ref{KdV_equation}) using the three-dimensional KP equation (\ref{KP_equation}).  
\section{Stability analysis}\label{sec:Stability analysis}
To analyze the stability of the solitary wave solution (\ref{phi1}) of the KP equation (\ref{KP_equation}), we use following transformation of the independent variables:
\begin{eqnarray}\label{transform variable}
X =\xi -U\tau , \eta^{\prime}=\eta , \zeta^{\prime}=\zeta, \tau^{\prime}=\tau.
\end{eqnarray}
Using the transformation (\ref{transform variable}), the KP equation (\ref{KP_equation}) can be written as
\begin{eqnarray}\label{transform form of KP}
\frac{\partial}{\partial X}\Big[-U\phi^{(1)}_{X}+\phi^{(1)}_{\tau}+ AB_{1}\phi^{(1)} \phi^{(1)}_{X}+ \frac{1}{2}A C \phi^{(1)}_{XXX}\Big] +\frac{1}{2}AD \Big(\phi^{(1)}_{\eta \eta}+\phi^{(1)}_{\zeta \zeta}\Big)=0,
\end{eqnarray}
where we drop the primes on the independent variables $\eta$, $\zeta$ and $\tau$ to simplify the notations. 

It is simple to check that the equation (\ref{transform form of KP}) is satisfied by the expression of $\phi^{(1)}(=\phi_{0}(X))$ as given in (\ref{phi1}). It is also important to note that $\displaystyle \phi^{(1)}_{\tau}=\frac{\partial \phi^{(1)}}{\partial \tau}=0$ for the expression of $\phi^{(1)}(=\phi_{0}(X))$ as given in (\ref{phi1}) and so this solution is the steady state solution of the KP equation (\ref{KP_equation}) or the corresponding KdV equation (\ref{KdV_equation}).

As $\phi_{0}(X)$ is a steady state solution of (\ref{transform form of KP}), we can decompse $\phi^{(1)}$ as
\begin{eqnarray}\label{decompose phi}
\phi^{(1)}=\phi_{0}(X)+ q(X, \eta, \zeta, \tau),
\end{eqnarray}
where $q(X, \eta, \zeta, \tau)$ is the perturbed part of $\phi^{(1)}$.

Substituting (\ref{decompose phi}) into (\ref{transform form of KP}) and then linearizing it with respect to $q$, we get the following linear equation for $q$:
\begin{eqnarray}\label{further_transform_form_of_KP}
\frac{\partial}{\partial X}\Big[-Uq_{X}+q_{\tau}+ AB_{1}(\phi_{0} q)_{X}+ \frac{1}{2}A C q_{XXX}\Big] +\frac{1}{2}AD \Big(q_{\eta \eta}+q_{\zeta \zeta}\Big)=0,
\end{eqnarray} 
where we have used the equation (\ref{phi1}) to simplify the above equation and and we have used the following notations: 
$\displaystyle q_{\tau} = \frac{\partial q}{\partial \tau}$, $\displaystyle q_{X} = \frac{\partial q}{\partial X}$, 
$\displaystyle q_{XXX} = \frac{\partial^{3} q}{\partial X^{3}}$, $\displaystyle (\phi_{0}q)_{X} = \frac{\partial }{\partial X}(\phi_{0}q)$.
 
Now, for long wave length plane wave perturbation along a direction  having direction cosines $l$, $m$, $n$, we take 
\begin{eqnarray}\label{exp of q}
q(X, \eta, \zeta, \tau) = \overline{q}(X)e^{i\{k(l X+m \eta+n \zeta)-\omega\tau\}},
\end{eqnarray}
where $k$ is small and $l^{2}+m^{2}+n^{2}=1$.

Due to the space time dependence of $q$ for long wave length plane wave perturbation as described in equation (\ref{exp of q}), the equation (\ref{further_transform_form_of_KP}) of $q$ transforms to the following equation of $\overline{q}$:
\begin{eqnarray}\label{equation_of_q_bar}
&& (M_{1}\overline{q})_{XX}-i \omega \overline{q}_{X}+kl\Big\{\omega \overline{q}+2i(M_{1}\overline{q})_{X}+iAC \overline{q}_{XXX}\Big\}\nonumber \\
&& - k^{2}l^{2}\Big\{M_{1}\overline{q}+\frac{5}{2}AC \overline{q}_{XX}+\frac{1}{2}AD\frac{m^{2}+n^{2}}{l^{2}}\overline{q}\Big\}\nonumber \\
&& -k^{3}l^{3}\Big\{2iAC\overline{q}_{X}\Big\}+k^{4}l^{4}\Big\{\frac{1}{2}AC\overline{q}\Big\}=0,
\end{eqnarray}
where
\begin{eqnarray}\label{M_equation}
M_{1}=-U + AB_{1} \phi_{0} + \frac{1}{2} AC \frac{\partial^{2}}{\partial X^{2}}.
\end{eqnarray}

Following the multiple-scale perturbation expansion method of Allen and Rowlands,\cite{allen1993determination,allen1995stability} we expand $\overline{q}(X)$ and $\omega$ as
\begin{eqnarray}\label{expansion of_{q}(X)}
\overline{q}(X) = \sum_{j=0}^{\infty}k^{j}q^{(j)}(X,X_{1},X_{2},X_{3},...),
\end{eqnarray}
\begin{eqnarray}\label{omega}
\omega = \sum_{j=0}^{\infty}k^{j}\omega^{(j)},
\end{eqnarray}
where $\omega^{(0)}=0$,  $X_{j}=k^{j}X$, $j=0,1,2,3...,$ and each $q^{(j)}(=q^{(j)}(X,X_{1},X_{2},X_{3},...))$ is a function of $X,X_{1},X_{2},X_{3},\cdots$ . It is important to note that $X_{0}=X$.

Finally, substituting (\ref{expansion of_{q}(X)}) and (\ref{omega}) into the equation (\ref{equation_of_q_bar}) and then equating the coefficients of different powers of $k$ on the both sides of the resulting equation, we get the following sequence of equations:
\begin{eqnarray}\label{sequence_of_eq}
\frac{\partial}{\partial X}(M_{1} q^{(j)})=Q^{(j)},
\end{eqnarray}
where
\begin{eqnarray}\label{Q^{j}_equation}
Q^{(j)}=\int_{\infty}^{X}{R^{(j)}} dX,
\end{eqnarray}
and $R^{(j)}$ for $j$ = 0, 1, 2, 3 are given in Appendix A.

Following Das \textit{et al.} \cite{das2014stability}, the general solution of (\ref{sequence_of_eq}) can be written in the following form:
\begin{eqnarray}\label{general solution}
q^{(j)}=A^{(j)}_{1}f+ A^{(j)}_{2}g+ A^{(j)}_{3}h+ \chi^{(j)},
\end{eqnarray}
where $A^{(j)}_{1}$, $A^{(j)}_{2}$, $A^{(j)}_{3}$ are all arbitrary functions of $X_{1},X_{2},X_{3},...$, and $f$, $g$, $h$, and, $\chi^{(j)}$ are given by
\begin{eqnarray}\label{fgh100}
f = \frac{d \phi_{0}}{dX},\label{f}
g = f\int\frac{1}{f^2}dX,\label{g}
h = f\int\frac{\phi_{0}}{f^{2}}dX,\label{h}
\end{eqnarray}
\begin{eqnarray}\label{chi_j}
\chi^{(j)} = \frac{2}{AC}  f\int\frac{\int \Big(f\int Q^{(j)}dX \Big)dX}{f^{2}}dX.
\end{eqnarray}
Using (\ref{fgh100}) and (\ref{chi_j}), we can write equation (\ref{general solution}) in the following form:
\begin{eqnarray}\label{reduce general sol}
q^{(j)}= A_{1}^{(j)}f- \frac{W^2 }{8 a} A_{2}^{(j)}S^{-2}-\frac{W^{2}}{16 a} \left(5 A_{2}^{(j)}+ 4 a A_{3}^{(j)}\right)+\frac{3 W^{2}}{16 a^{2}}  \left(5 A_{2}^{(j)}+ 4 a A_{3}^{(j)}\right)\phi_{0}\nonumber \\+\frac{3 W^{2}}{32 a^{2}} \left(5 A_{2}^{(j)}+ 4 a A_{3}^{(j)}\right)f X +\frac{2}{AC}  f\int\frac{\int \Big(f\int Q^{(j)}dX \Big)dX}{f^{2}}dX.
\end{eqnarray}
Here $S=\mbox{sech}[\frac{X}{W}]$ and $\phi_{0}$ is given by (\ref{phi1}).

\subsection{Zeroth order equation}
The solution (\ref{reduce general sol}) of the equation (\ref{sequence_of_eq}) for $j=0$ can be written as 
\begin{eqnarray}\label{general solution for j=0}
q^{(0)} &=& A_{1}^{(0)}f- \frac{W^2 }{8 a} A_{2}^{(0)}S^{-2}- \frac{W^{2}}{16 a} \left(5 A_{2}^{(0)}+ 4 a A_{3}^{(0)}\right)\nonumber \\
&& +\frac{3 W^{2}}{16 a^{2}}  \left(5 A_{2}^{(0)}+ 4 a A_{3}^{(0)}\right)\phi_{0}+\frac{3 W^{2}}{32 a^{2}} \left(5 A_{2}^{(0)}+ 4 a A_{3}^{(0)}\right)f X,
\end{eqnarray}
where we have used the equation (\ref{R0}) of Appendix A and the equation (\ref{Q^{j}_equation}) to find $Q^{(0)}$.

Now it is simple to check that $S^{-2}$ $\rightarrow +\infty$ as $X \rightarrow +\infty$. Therefore, to make $q^{(0)}$ bounded we must have 
\begin{eqnarray}\label{A2}
- \frac{W^2 }{8 a} A_{2}^{(0)}=0 \Leftrightarrow A_{2}^{(0)}=0.
\end{eqnarray}
Using (\ref{A2}), the equation (\ref{general solution for j=0}) can be written in the following form:
\begin{eqnarray}\label{general solution for j=0_100}
q^{(0)} &=& A_{1}^{(0)}f- \frac{W^{2}}{4} A_{3}^{(0)} +\frac{3 W^{2}}{4 a}  A_{3}^{(0)}\phi_{0}+\frac{3 W^{2}}{8 a} A_{3}^{(0)}f X.
\end{eqnarray}
Again, $q^{(0)}$ is consistent at $X = +\infty$, i.e., $q^{(0)}$ $\rightarrow 0 $ as $X \rightarrow +\infty$, if we choose 
\begin{eqnarray}\label{A3}
- \frac{W^{2}}{4} A_{3}^{(0)}=0 \Leftrightarrow  A_{3}^{(0)} =0.
\end{eqnarray}
Therefore, the equation (\ref{general solution for j=0_100}) takes the following form:
\begin{eqnarray}\label{zero order sol}
q^{(0)}=A^{(0)}_{1}f.
\end{eqnarray}

\subsection{First order equation}
Using the equation (\ref{R1}) of Appendix A and (\ref{zero order sol}), the solution (\ref{reduce general sol}) of the differential equation (\ref{sequence_of_eq}) for $j=1$ can be put in the following form:
\begin{eqnarray}\label{general solution for j=1}
q^{(1)} &=& A_{1}^{(1)}f- \frac{W^2 }{8 a} A_{2}^{(1)}S^{-2}-\frac{W^{2}}{16 a} \left(5 A_{2}^{(1)}+ 4 a A_{3}^{(1)}\right)\nonumber\\
&&+\Big\{\frac{3 W^{2}}{16 a^{2}}  \left(5 A_{2}^{(1)}+ 4 a A_{3}^{(1)}\right)+ iA^{(0)}_{1} \frac{ \omega^{(1)}}{ U} \Big\} \phi_{0}\nonumber\\
&&+\Big\{- \frac{\partial A^{(0)}_{1}}{\partial X_{1}}+ iA^{(0)}_{1} \frac{\omega^{(1)}- 2lU }{2U} +\frac{3 W^{2}}{32 a^{2}} \left(5 A_{2}^{(1)}+ 4 a A_{3}^{(1)}\right)\Big\}fX,
\end{eqnarray}
where we have used MATHEMATICA \cite{wolfram1999mathematica} to compute all the integrals of right hand side of (\ref{reduce general sol}) for $j=1$.

Now $q^{(1)}$ can be made bounded and consistent at $X=+\infty$ if and only if $A^{(1)}_{2}=A^{(1)}_{3}=0$ and consequently the equation (\ref{general solution for j=1}) can be written in the form:
\begin{eqnarray}\label{reduce 1st order sol}
q^{(1)}=A_{1}^{(1)}f+ iA^{(0)}_{1} \frac{ \omega^{(1)}}{ U} \phi_{0}- \bigg\{\frac{\partial A^{(0)}_{1}}{\partial X_{1}}- iA^{(0)}_{1} \frac{\omega^{(1)}- 2lU }{2U}\bigg\}fX,
\end{eqnarray}
where we have used exactly the same argument as given in the lowest order equation to get the bounded and consistent solution $q^{(0)}$.
 
Following Allen and Rowlands, \cite{allen1993determination,allen1995stability} the first term on the right hand side of (\ref{reduce 1st order sol}) can be removed because this type of term has already been included in $q^{(0)}$. Again, according to the prescription of Allen and Rowlands, \cite{allen1993determination,allen1995stability} the last term on the right hand side of (\ref{reduce 1st order sol}) is a ghost secular term and this term can be removed from the equation (\ref{reduce 1st order sol}) if we choose
\begin{eqnarray}\label{A01_X1}
\frac{\partial A^{(0)}_{1}}{\partial X_{1}}=iA^{(0)}_{1} \frac{\omega^{(1)}- 2lU }{2U}.
\end{eqnarray}
Therefore, the equation (\ref{reduce 1st order sol}) can be put in the following form: 
\begin{eqnarray}\label{q1_final}
q^{(1)}= iA^{(0)}_{1} \frac{ \omega^{(1)}}{ U} \phi_{0}.
\end{eqnarray}

\subsection{Second order equation}
Using the equation (\ref{R2}) of Appendix A, (\ref{zero order sol}) and (\ref{q1_final}), the solution (\ref{reduce general sol}) of the equation (\ref{sequence_of_eq}) for $j=2$ can be written as   
\begin{eqnarray}\label{general solution for j=2}
q^{(2)} &=& - \frac{W^2 }{8 a} A_{2}^{(2)}S^{-2} - \frac{W^{2}}{16a}\left(5 A_{2}^{(2)}+ 4 a A_{3}^{(2)} \right)
- a A_{1}^{(0)}\frac{W}{4U^{2}}(\omega^{(1)})^{2} R\nonumber \\
&&- a A_{1}^{(0)}\frac{W}{12U^{2}} \Big\{3(\omega^{(1)})^{2}- 2 U V (m^{2}+n^{2})\Big\}RS^{-2} \nonumber\\
&&+\bigg\{\frac{3 W^{2}}{16 a^{2}}  \left(5 A_{2}^{(2)}+ 4 a A_{3}^{(2)}\right)+ iA^{(0)}_{1} \frac{ \omega^{(2)}}{ U} \bigg\} \phi_{0}\nonumber\\
&&+\Big\{\frac{3 W^{2}}{32 a^{2}} \left(5 A_{2}^{(2)}+ 4 a A_{3}^{(2)}\right)+ iA^{(0)}_{1} \frac{\omega^{(2)}}{2U}- \frac{\partial A^{(0)}_{1}}{\partial X_{2}}\Big\}fX ,
\end{eqnarray}
where $R=\mbox{tanh}[\frac{X}{W}]$, we have used MATHEMATICA \cite{wolfram1999mathematica} to find all the integrals of (\ref{reduce general sol}) for $j=2$ and we have ignored the terms that have been included in the zeroth and first order solutions $q^{(0)}$ and $q^{(1)}$. 

The above expression of $q^{(2)}$ shows that $q^{(2)}$ is not bounded because of the presence of the exponential secular terms $\frac{1}{S^{2}}$ and $\frac{R}{S^{2}}$. So, it is impotant to remove both the exponential secular terms simultaneously. Now, to remove both the exponential secular terms simultaneously, it is necessary to make a common factor of the coefficients of these two terms equal to zero. As $W \neq 0$, to get a common factor between the coefficients of the two exponential secular terms in the expression of $q^{(2)}$, we write the arbitrary function $A^{(2)}_{2}$ in the following form:
\begin{eqnarray}\label{A22}
A^{(2)}_{2} = B^{(2)}_{2} \Big\{3(\omega^{(1)})^{2}- 2 U V (m^{2}+n^{2})\Big\},
\end{eqnarray}
and consequently $q^{(2)}$ can be written as
\begin{eqnarray}\label{general solution for j=2_100}
q^{(2)} &=&  - \frac{W^{2}}{16a}\left(5 A_{2}^{(2)}+ 4 a A_{3}^{(2)} \right)
- a A_{1}^{(0)}\frac{W}{4U^{2}}(\omega^{(1)})^{2} R\nonumber \\
&& - \Big\{3(\omega^{(1)})^{2}- 2 U V (m^{2}+n^{2})\Big\}\Big\{B_{2}^{(2)}\frac{W^2 }{8 a} \frac{1}{S^{2}}+  A_{1}^{(0)}\frac{aW}{12U^{2}} \frac{R}{S^{2}}\Big\} \nonumber\\
&&+\bigg\{\frac{3 W^{2}}{16 a^{2}}  \left(5 A_{2}^{(2)}+ 4 a A_{3}^{(2)}\right)+ iA^{(0)}_{1} \frac{ \omega^{(2)}}{ U} \bigg\} \phi_{0}\nonumber\\
&&+\Big\{\frac{3 W^{2}}{32 a^{2}} \left(5 A_{2}^{(2)}+ 4 a A_{3}^{(2)}\right)+  iA^{(0)}_{1} \frac{\omega^{(2)}}{2U}- \frac{\partial A^{(0)}_{1}}{\partial X_{2}}\Big\}fX ,
\end{eqnarray}
where $B^{(2)}_{2}$ is another arbitrary functions of $X_{1},X_{2},X_{3},\cdots$.

From the expression of $q^{(2)}$ as given in (\ref{general solution for j=2_100}), we find that exponential secularity in the resulting expression of $q^{(2)}$ is due to the presence of the terms $S^{-2}$ and $RS^{-2}$. Therefore, we can remove the exponential secular terms from the expression of $q^{(2)}$ by setting 
\begin{eqnarray}\label{cc1}
3(\omega^{(1)})^{2}- 2 U V (m^{2}+n^{2})=0.
\end{eqnarray}
This equation gives the following expression of $(\omega^{(1)})^{2}$ :
\begin{eqnarray}\label{w_1_2}
(\omega^{(1)})^{2}= \frac{2}{3} U V (m^{2}+n^{2}).
\end{eqnarray}
The above equation is exactly same as the equation (47) of Sardar \textit{et al.} \cite{sardar2016stability} for $r=1$. But they have used the small-$k$ perturbation expansion method of Rowlands and Infeld \cite{rowlands1969stability,infeld1972stability,infeld1973stability,infeld1985self,zakharov1974instability} to investigate the lowest order stability of KP and different modified KP equations.

From equations (\ref{A22}) and (\ref{cc1}), we get $A_{2}^{(2)}=0$. To make $q^{(2)}$ consistent with the condition that $q^{(2)}$ $\rightarrow 0 $ as $X\rightarrow +\infty$, we have
\begin{eqnarray}\label{a23}
A_{3}^{(2)} &=& - a A_{1}^{(0)} \frac{(\omega^{(1)})^{2}}{WU^{2}}
\end{eqnarray} 
and consequently $q^{(2)}$ can be written as 
\begin{eqnarray}\label{q2}
q^{(2)} &=&  A_{1}^{(0)} \Big\{a  \frac{W}{4U^{2}}(\omega^{(1)})^{2} (1- R)+  t_{1}\phi_{0}\Big\}  +  \bigg\{\frac{1}{2} A^{(0)}_{1} t_{1}- \frac{\partial A^{(0)}_{1}}{\partial X_{2}} \bigg\}f X,
\end{eqnarray}
where
\begin{eqnarray}
t_{1}=\frac{i\omega^{(2)}}{U}-  \frac{3W}{4U^{2}}(\omega^{(1)})^{2} . \label{t1}
\end{eqnarray}
Again, according to the prescription of Allen and Rowlands, \cite{allen1993determination,allen1995stability} the last term in the above expression of $q^{(2)}$ is the ghost secular term and this term can be removed from the equation (\ref{q2}) if we set
\begin{eqnarray}
\frac{\partial A^{(0)}_{1}}{\partial X_{2}}= \frac{1}{2} A^{(0)}_{1} t_{1}.
\end{eqnarray}
 Therefore, the final form of $q^{(2)}$ can be written as follows:
\begin{eqnarray}\label{q2_final}
q^{(2)}= A^{(0)}_{1}\bigg[a \frac{W}{4U^{2}}(\omega^{(1)})^{2} (1- R)+ t_{1}\phi_{0} \bigg].
\end{eqnarray}
From the above expression of $q^{(2)}$, we see that $q^{(2)}$ is bounded for any real $X$ and $q^{(2)}$ is consistent at $X=+\infty$ but $q^{(2)}$ is not consistent at $X=-\infty$, i.e., $\displaystyle \lim_{X \rightarrow -\infty}q^{(2)} \neq 0$, but as $q^{(2)}$ is bounded, this inconsistency can be removed by proper grouping with the terms of higher orders. So, we consider the solution (\ref{reduce general sol}) for $j=3$.   
\subsection{Third order equation}
Using the equation (\ref{R3}) of Appendix A, (\ref{zero order sol}), (\ref{q1_final}) and (\ref{q2_final}), the solution (\ref{reduce general sol}) of the equation (\ref{sequence_of_eq}) for $j=3$ can be written as   
\begin{eqnarray}\label{general solution for j=3}
q^{(3)} &=& - \frac{W^2 }{8 a} A_{2}^{(3)}S^{-2}- \frac{W^{2}}{16a}\left(5 A_{2}^{(3)}+ 4 a A_{3}^{(3)} \right)\nonumber\\
&& +\frac{3 W^{2}}{16 a^{2}}  \left(5 A_{2}^{(3)}+ 4 a A_{3}^{(3)}\right) \phi_{0}+ \frac{3 W^{2}}{32 a^{2}} \left(5 A_{2}^{(3)}+ 4 a A_{3}^{(3)}\right)fX 
\nonumber\\
&&+ i A_{1}^{(0)}\frac{aW}{8U^{3}}\omega^{(1)} \bigg\{-2UVW(m^{2}+n^{2})+ W (\omega^{(1)})^{2} +  4iU\omega^{(2)}
\bigg\}  \frac{R}{S^{2}}\nonumber\\
&&+ i A_{1}^{(0)}\frac{aW}{4U^{3}}\omega^{(1)} \bigg\{-3UVW(m^{2}+n^{2})+ 3W(\omega^{(1)})^{2} +  2iU\omega^{(2)}
\bigg\} R \nonumber\\
&&+ i A_{1}^{(0)}\frac{aW}{8U^{3}}\omega^{(1)} \bigg\{-2UV(m^{2}+n^{2})+ 3(\omega^{(1)})^{2} \bigg\} \bigg\{-2X+ 2W Log\frac{1}{S} \bigg\} \nonumber\\
&&+ i A_{1}^{(0)}\frac{W}{4U^{3}}\omega^{(1)} \bigg\{2UV(m^{2}+n^{2})- 3(\omega^{(1)})^{2} \bigg\} \bigg\{-3X+ 3W Log\frac{1}{S} \bigg\} \phi_{0} \nonumber\\ 
&&+ i A_{1}^{(0)}\frac{1}{8U^{3}} \bigg[ W^{2} \bigg\{-8l^{3}U^{3}+ 4UV(m^{2}+n^{2})\omega^{(1)}- 5(\omega^{(1)})^{3} \bigg\}+ 8 U^{2} \omega^{(3)} \bigg] \phi_{0}\nonumber\\ 
&&+ i A_{1}^{(0)}\frac{3W^{3}}{16U^{3}}\omega^{(1)} \bigg\{2UV(m^{2}+n^{2})- 3(\omega^{(1)})^{2} \bigg\} PolyLog[2,-e^{-\frac{2X}{W}}]f \nonumber\\
&&+ i A_{1}^{(0)}\frac{a}{U^{3}W}\omega^{(1)} \bigg\{-2UV(m^{2}+n^{2})+ 3(\omega^{(1)})^{2} \bigg\}\nonumber\\
&&\times \bigg[\frac{48W^{3}}{128a} \bigg\{ Log[1+ e^{-\frac{2X}{W}}]-Log\frac{1}{S}\bigg\}fX+ 48 W^{2} fX^{2} \bigg]  \nonumber\\
&&+ \bigg[i A_{1}^{(0)}\frac{W^{2} \bigg\{-8l^{3}U^{3}+ 16UV(m^{2}+n^{2})\omega^{(1)}- 23(\omega^{(1)})^{3} \bigg\}+ 8 U^{2} \omega^{(3)}}{16U^{3}}\nonumber \\
&&- \frac{\partial A^{(0)}_{1}}{\partial X_{3}}\bigg]fX ,
\end{eqnarray}
where we have used MATHEMATICA \cite{wolfram1999mathematica} to find all the integrals of (\ref{reduce general sol}) for $j=3$ and ignored the terms already taken in lower order solutions. 

Substituting the expression of $(\omega^{(1)})^{2}$ as given in the equation (\ref{w_1_2}), the expression of $q^{(3)}$ can be simplified as follows:
\begin{eqnarray}\label{general solution for j=3_simplify}
q^{(3)} &=& - \frac{W^2 }{8 a} A_{2}^{(3)}S^{-2}- \frac{W^{2}}{16a}\left(5 A_{2}^{(3)}+ 4 a A_{3}^{(3)} \right)+\frac{3 W^{2}}{16 a^{2}}  \left(5 A_{2}^{(3)}+ 4 a A_{3}^{(3)}\right) \phi_{0}\nonumber\\
&&+ \frac{3 W^{2}}{32 a^{2}} \left(5 A_{2}^{(3)}+ 4 a A_{3}^{(3)}\right)fX- i A_{1}^{(0)}\frac{aW}{4U^{2}}\omega^{(1)} \bigg\{VW(m^{2}+n^{2})- 2i\omega^{(2)} \bigg\}  R\nonumber\\
&& - i A_{1}^{(0)}\frac{aW}{6U^{2}}\omega^{(1)} \bigg\{VW(m^{2}+n^{2})- 3i\omega^{(2)} \bigg\}  \frac{R}{S^{2}}\nonumber\\
&&+ i A_{1}^{(0)}\frac{1}{12U^{2}} \bigg\{-12l^{3}U^{2}W^{2}+ VW^{2}(m^{2}+n^{2})\omega^{(1)}+ 12U \omega^{(3)} \bigg\} \phi_{0} \nonumber\\
&&+ \bigg[i A_{1}^{(0)}\frac{1}{24U^{2}} \bigg\{-12l^{3}U^{2}W^{2}+ VW^{2}(m^{2}+n^{2})\omega^{(1)}+ 12U \omega^{(3)} \bigg\}- \frac{\partial A^{(0)}_{1}}{\partial X_{3}}\bigg]fX,\nonumber\\ 
\end{eqnarray}
From the expression of $q^{(3)}$ as given in (\ref{general solution for j=3_simplify}), we see that exponential secular terms in the expression of $q^{(3)}$ is due to the terms $\frac{1}{S^{2}}$ and $\frac{R}{S^{2}}$. To remove both the exponential secular terms simultaneously, it is necessary to make a common factor of the coefficients of these two  terms equal to zero. As $W \neq 0$, to get a common factor of the coefficients of the two exponential secular terms in the expression of $q^{(3)}$, we set  
\begin{eqnarray}\label{q^{(3)}_bdd}
A_{2}^{(3)}= B_{2}^{(3)} \bigg\{VW(m^{2}+n^{2})- 3i\omega^{(2)} \bigg\},
\end{eqnarray}
where $B_{2}^{(3)}$ is another arbitrary functions of $X_{1},X_{2},X_{3},...$.

Therefore, the equation (\ref{general solution for j=3_simplify}) assumes the following form:
\begin{eqnarray}\label{general solution for j=3_simplify100}
q^{(3)} &=& - \frac{W^{2}}{16a}\left(5 A_{2}^{(3)}+ 4 a A_{3}^{(3)} \right)+\frac{3 W^{2}}{16 a^{2}}  \left(5 A_{2}^{(3)}+ 4 a A_{3}^{(3)}\right) \phi_{0} \nonumber \\ && + \frac{3 W^{2}}{32 a^{2}} \left(5 A_{2}^{(3)}+ 4 a A_{3}^{(3)}\right)fX- i A_{1}^{(0)}\frac{aW}{4U^{2}}\omega^{(1)} \bigg\{VW(m^{2}+n^{2})- 2i\omega^{(2)} \bigg\}  R\nonumber\\
&& - \bigg\{VW(m^{2}+n^{2})- 3i\omega^{(2)} \bigg\} \bigg\{\frac{W^2 }{8 a} B_{2}^{(3)}\frac{1}{S^{2}}+ i A_{1}^{(0)}\frac{aW}{6U^{2}}\omega^{(1)}  \frac{R}{S^{2}}\bigg\}\nonumber\\
&&+ i A_{1}^{(0)}\frac{1}{12U^{2}} \bigg\{-12l^{3}U^{2}W^{2}+ VW^{2}(m^{2}+n^{2})\omega^{(1)}+ 12U \omega^{(3)} \bigg\} \phi_{0} \nonumber\\
&&+ \bigg[i A_{1}^{(0)}\frac{1}{24U^{2}} \bigg\{-12l^{3}U^{2}W^{2}+ VW^{2}(m^{2}+n^{2})\omega^{(1)}+ 12U \omega^{(3)} \bigg\}- \frac{\partial A^{(0)}_{1}}{\partial X_{3}}\bigg]fX,\nonumber\\ 
\end{eqnarray}
We can remove the exponential secular terms from the expression of $q^{(3)}$ by setting 
\begin{eqnarray}\label{cc}
VW(m^{2}+n^{2})- 3i\omega^{(2)} =0.
\end{eqnarray}
From equations (\ref{q^{(3)}_bdd}) and (\ref{cc}), we get $A_{2}^{(3)}=0$. Using this value of $A_{2}^{(3)}$ and the condition that $q^{(3)}$ is consistent at $X=+\infty$, we have
 \begin{eqnarray}\label{q^{(3)}_consistency}
A_{3}^{(3)}= - i A_{1}^{(0)}\frac{a}{WU^{2}}\omega^{(1)} \bigg\{VW(m^{2}+n^{2})- 2i\omega^{(2)} \bigg\}.
 \end{eqnarray}
 Now $q^{(3)}$ can be written as follows:
 \begin{eqnarray}\label{general solution for j=3_simplify_1}
q^{(3)}=  i A_{1}^{(0)}\frac{aW}{4U^{2}}\omega^{(1)} \bigg\{VW(m^{2}+n^{2})- 2i\omega^{(2)} \bigg\}  (1-R)  + i A_{1}^{(0)}r_{1} \phi_{0},
\end{eqnarray}
where
\begin{eqnarray}\label{r1}
r_{1}=\frac{1}{6U^{2}} \bigg\{-6l^{3}U^{2}W^{2}+ W\omega^{(1)}\{-4VW (m^{2}+n^{2})+9i \omega^{(2)}\}+ 6U \omega^{(3)} \bigg\},
\end{eqnarray}
and we have used the following equation to remove the ghost secular term appearing in the expression of $q^{(3)}$
\begin{eqnarray}\label{r11}
\frac{\partial A^{(0)}_{1}}{\partial X_{3}}=\frac{1}{2} i A_{1}^{(0)} r_{1}.
\end{eqnarray} 
The above expression of $q^{(3)}$ shows that $q^{(3)}$ is bounded for any real $X$ and $q^{(3)}$ is consistent at $X=+\infty$ but $q^{(3)}$ is not consistent at $X=-\infty$, i.e., $\displaystyle \lim_{X \rightarrow -\infty}q^{(3)} \neq 0$. Again, it is important to note that $q^{(3)}$ contains same type of terms of $q^{(2)}$, viz., $(1-R) $ and $\phi_{0}$.  As $q^{(2)}$ and $q^{(3)}$ both are bounded, the inconsistent term of $q^{(2)}$ and $q^{(3)}$ can be removed by proper grouping with the higher order terms.

The equation (\ref{cc}) gives the following expression for $\omega^{(2)}$:
\begin{eqnarray}\label{second_order_1}
\omega^{(2)}=-\frac{i}{3} VW (m^{2}+n^{2}).
\end{eqnarray}
Equation (\ref{second_order_1}) shows that $\omega^{(2)}$ is imaginary along with $i\omega^{(2)}>0$ and consequently, the solitary wave solution (\ref{phi1}) is stable at the order $k^{2}$.

\section{Conclusions}\label{sec:conclusions}
In the present paper, we have seen that solitary wave solution of the KdV equation is stable up to the order $k^{2}$, where $k$ is the wave number for long-wavelength plane-wave perturbation. From the physical point of view, this problem gives an idea of finding nonlinear dispersion relation between $\omega$ and $k$ when the solitary wave solution (\ref{phi1}) is the steady state solution of the nonlinear evolution equation (\ref{KP_equation}) for small value of the wave number $k$.

\noindent \textbf{APPENDIX A}: $R^{(j)}$ - Integrand of the integration in equation (\ref{Q^{j}_equation}) for $j$ = 0, 1 , 2 and 3.
\begin{eqnarray}
R^{(0)} &=& 0,\label{R0}\\
R^{(1)} &=& i\omega^{(1)}q^{(0)}_{0}- 2il[M_{1}q^{(0)}]_{0}- iAClq^{(0)}_{000}\nonumber\\
&&- ACq^{(0)}_{0001}- 2 [M_{1}q^{(0)}]_{01},\label{R1}\\
R^{(2)} &=& \bigg[\frac{1}{2}AD(m^{2}+n^{2})- l\omega^{(1)}\bigg]q^{(0)}+ l^{2}M_{1}q^{(0)}\nonumber\\
&&+ i \omega^{(2)} q^{(0)}_{0}+ i \omega^{(1)}[q^{(0)}_{1}+ q^{(1)}_{0}] \nonumber\\
&&- i2l[(M_{1}q^{(0)})_{1}+ (M_{1}q^{(1)})_{0}]\nonumber\\
&&+ \frac{5}{2}ACl^{2}q^{(0)}_{00} - iACl[5q^{(0)}_{001}+ q^{(1)}_{000}]\nonumber\\
&&- \frac{1}{2}AC[2q^{(0)}_{0002}+ 5q^{(0)}_{0011}+ 2q^{(1)}_{0001}]- [M_{1}q^{(0)}]_{11}\nonumber\\
&&- 2 [M_{1}q^{(0)}]_{02}- 2[M_{1}q^{(1)}]_{01},\label{R2}
\end{eqnarray}
\begin{eqnarray}
R^{(3)} &=& \bigg[\frac{1}{2}AD(m^{2}+n^{2})- l\omega^{(1)}\bigg]q^{(1)}+ l^{2}M_{1}q^{(1)}\nonumber\\
&&+ i \omega^{(3)} q^{(0)}_{0}+  \omega^{(2)}[i(q^{(0)}_{1}+ q^{(1)}_{0})- lq^{(0)}] \nonumber\\
&&+ i\omega^{(1)}[q^{(0)}_{2}+ q^{(1)}_{1}+ q^{(2)}_{0}]- i2l[(M_{1}q^{(0)})_{2}\nonumber\\
&&+ (M_{1}q^{(1)})_{1} + (M_{1}q^{(2)})_{0}]+ ACl^{2}[6q^{(0)}_{01} + \frac{5}{2}q^{(1)}_{00} ]\nonumber\\
&&- iACl[5q^{(0)}_{002}+ 6q^{(0)}_{011}+ 5 q^{(1)}_{001}+ q^{(2)}_{000}]\nonumber\\
&&- \frac{1}{2}AC [4q^{(0)}_{0111}+ 10q^{(0)}_{0012}+ 2q^{(0)}_{0003}+ 2q^{(1)}_{0002}\nonumber\\
&&+ 5q^{(1)}_{0011}+ 2q^{(2)}_{0001}]- 2[M_{1}q^{(0)}]_{03}- 2[M_{1}q^{(0)}]_{12}\nonumber\\
&&- 2[M_{1}q^{(1)}]_{02}- [M_{1}q^{(1)}]_{11}- 2[M_{1}q^{(2)}]_{01},\label{R3}
\end{eqnarray}
where
\begin{eqnarray}\label{qjr}
q^{(j)}_{r} = \frac{\partial q^{(j)}}{\partial X_{r}},~
q^{(j)}_{rs} = \frac{\partial^{2} q^{(j)}}{\partial X_{r}\partial X_{s}},~
q^{(j)}_{rst} = \frac{\partial^{3} q^{(j)}}{\partial X_{r}\partial X_{s}\partial X_{t}}\nonumber
\end{eqnarray}
\begin{eqnarray}\label{qjry}
q^{(j)}_{rsty} = \frac{\partial^{4} q^{(j)}}{\partial X_{r}\partial X_{s}\partial X_{t}\partial X_{y}},~
[M_{1}q^{(j)}]_{rs}=\frac{\partial^{2} (M_{1}q^{(j)})}{\partial X_{r}\partial X_{s}}\nonumber
\end{eqnarray}

\providecommand{\noopsort}[1]{}\providecommand{\singleletter}[1]{#1}%

\end{document}